%% file: spring-networks-letter.tex
\documentclass[%
reprint,
superscriptaddress,
 amsmath,amssymb,
aps,
prl,
]{revtex4-1}

\input{shortcuts}
\usepackage{graphicx,xcolor}
\usepackage{dcolumn}
\usepackage{bm}
\usepackage[normalem]{ulem}
\usepackage{bbm}
\usepackage{hyperref}
\usepackage[capitalise]{cleveref}
\usepackage{siunitx}
\usepackage{mathtools}
\usepackage[utf8]{inputenc}

\usepackage{amsfonts}
\usepackage{amstext}
\usepackage{amsmath}
\usepackage{amsthm}
\usepackage{amssymb}
\usepackage{amsbsy}   

\newcommand{\sn}[1]{{\color{black}{#1}}}

\newcommand{\mw}[1]{{\color{black}{#1}}}

\newcommand{\lbar}[0]{{\boldsymbol{\bar l}}}
\newcommand{\barz}[0]{z}
\newcommand{\lstar}[0]{\boldsymbol{l^*}}
\newcommand{\ellstar}[0]{{\ell^*}}

\newcommand{\ellbar}[0]{{\bar \ell}}

\newcommand{\dl}[0]{\boldsymbol{\Delta l}}
\newcommand{\dell}[0]{{\Delta \ell}}
\newcommand{\dlgiven}[0]{{\dl|\lbar=\bar \ell}}

\newcommand{\varlbar}[0]{{\Var(\bar l)}}
\newcommand{\vardlgiven}[0]{{\Var(\dl|\lbar)}}
\newcommand{\vardlgivenl}[0]{{\Var(\dl|\lbar=\bar\ell)}}
\newcommand{\meanvar}[0]{{\E_{\lbar}[\vardlgiven]}}

\hypersetup{pdfauthor={Knut Heidemann}, pdftitle={Topology determines force distributions in one-dimensional random spring networks},pdfkeywords={force distribution, topology, random graphs, spring network}}

\let\emph\relax 
\DeclareTextFontCommand{\emph}{\textnormal\em}

\begin{document}

\title{Topology counts: force distributions in circular spring networks}

\author{Knut M. Heidemann}
\affiliation{Institute for Numerical and Applied Mathematics, University of Goettingen, Germany}
\author{Andrew O. Sageman-Furnas}
\affiliation{Institute for Numerical and Applied Mathematics, University of Goettingen, Germany}
\author{Abhinav Sharma}%
\affiliation{Third Institute of Physics---Biophysics, University of Goettingen, Germany}%

\author{Florian Rehfeldt}
\affiliation{Third Institute of Physics---Biophysics, University of Goettingen, Germany}%

\author{Christoph F. Schmidt}
\email{cfs@physik3.gwdg.de}
\affiliation{Third Institute of Physics---Biophysics, University of Goettingen, Germany}%

\author{Max Wardetzky}
\email{wardetzky@math.uni-goettingen.de}
\affiliation{Institute for Numerical and Applied Mathematics, University of Goettingen, Germany}

\date{\today}


\begin{abstract}
Filamentous polymer networks govern the mechanical properties of many biological materials.
Force distributions within these networks are typically highly inhomogeneous and, although the importance of force distributions for structural properties is well recognized, they are far from being understood quantitatively.
Using a combination of \emph{probabilistic} and \emph{graph-theoretical} techniques we derive force distributions in a model system consisting of ensembles of \emph{random} linear spring networks on a circle.
We show that characteristic quantities, such as mean and variance of the force supported by individual springs, can be derived explicitly in terms of only two parameters: (i) \emph{average connectivity} and (ii) \emph{number of nodes}.
Our analysis shows that a classical mean-field approach \emph{fails} to capture these characteristic quantities correctly.
In contrast, we demonstrate that \emph{network topology} is a crucial determinant of force distributions in an elastic spring network.
\end{abstract}
\maketitle

\graphicspath{{img/}}


Filamentous polymer networks are ubiquitous in nature. They make up the cytoskeleton of animal cells and form the scaffold of the extracellular matrix in, e.g., connective tissue.
\sn{These networks determine the mechanical response of cells and tissues and support elastic forces under external or internal loading at both mesoscopic and macroscopic scales.}
The force distributions within such networks can be highly inhomogeneous \cite{heussinger2007force,arevalo2015stress}.
Internal forces in these typically non-equilibrium networks result mostly from molecular motors \cite{mizuno2007nonequilibrium,koenderink2009active,alvarado2013molecular} in the cell cytoskeleton or, on a larger scale, from cells embedded in extracellular matrices, such as platelets in blood clots \cite{shah1997strain,lam2011mechanics}.


The quantitative analysis of force distributions within random polymer networks has largely relied on computational modeling \cite{heussinger2007force,Heidemann2014}.
Analytical descriptions of filamentous networks have primarily used \emph{effective-medium} \cite{Thorpe1985,Broedersz2011b,Sheinman2012a} or \emph{mean-field} \cite{Storm2005,Sharma2013a,Heidemann2014} approaches.
Effective-medium theories rely on mapping a disordered system to an ordered one.
It is unclear, however, how force distributions change under this mapping.
Mean-field approaches do not consider the full network topology, but only the \emph{local} degree of connectivity. 
We show that such an approach fails to describe force distributions even for a very simple model system; 
\sn{in fact, topological features, i.e., cycles/loops in the networks, cause global coupling that remains prevalent even when the system becomes large.}

\begin{figure}
  \centering
  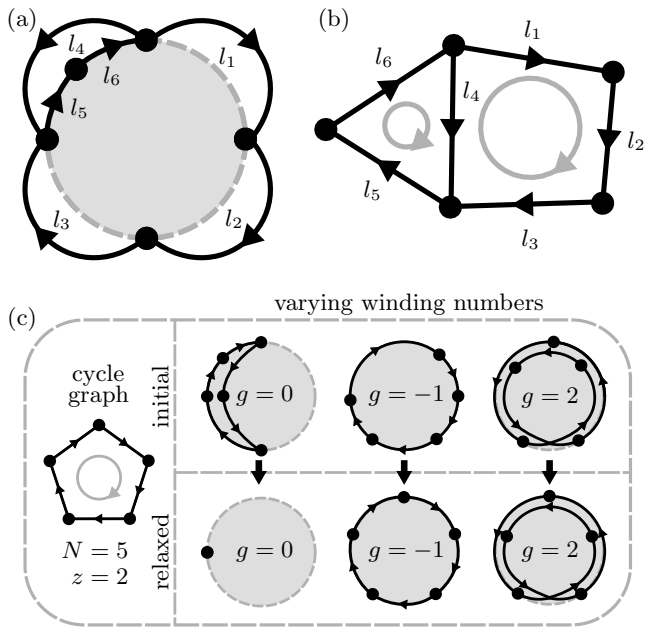
  \caption{(a) An example network on the circle, with $N=\num{5}$ and $z=\num{2.4}$. (b) Graph representation of the network in (a). The edge/spring orientations are depicted by black arrows. The network contains two fundamental cycles, for example: $\{l_1,l_2,l_3,l_4\}$ and $\{l_4,l_5,l_6\}$. After choosing arbitrary orientations for both cycles (gray arrows), we construct linear constraints that fix their winding numbers (\cref{eq:totalEnergy})---here: $l_1+l_2+l_3-l_4 = -1$ (winds around circle once) and $l_4+l_5+l_6 = 0$ (contractible). \sn{(c) The abstract cycle graph ($z=2$) with $N=5$ (left) and three realizations on the circle with distinct topologies (same graphs but different winding numbers $g$). Top and bottom row show initial and corresponding relaxed configurations, respectively. Note that, for visualization purposes, overlapping springs are drawn with a slight offset.}}
  \label{img:schematic}
\end{figure}

The simple model system that we consider here consists of ensembles of one-dimensional random spring networks on a circle.
Considering such networks is equivalent to applying \emph{periodic boundary conditions} in one dimension.
To model a generically forced system we employ a generation procedure that results in initial configurations that are not in mechanical equilibrium. We then study the resulting force distributions of the relaxed systems.

We generate \emph{initial} network configurations as follows (\cref{img:schematic}):
(i) Place $N$ node positions \sn{(indexed from 1 to $N$)} drawn from a \emph{uniform} distribution on the circle.
(ii) Connect these nodes in the order given by their indices into one connected \emph{cycle} via springs.
We always connect consecutive nodes via the \emph{shorter} of the two possible distances.
\sn{Note that the cycle may wrap around the circle zero, one, or multiple times (\cref{img:schematic}~(c)).}
This step guarantees that each network will always have only one connected component \sn{and prevents dangling ends}.
(iii) Connect further node pairs \emph{randomly}, such that each node pair is connected by at most one spring, until the network contains $Nz/2$ springs, where the \emph{average degree} of \sn{connectivity} $\barz$ is chosen such that $Nz/2$ is an integer.

Each spring is linear, has rest length zero, and unit spring constant. Its length is measured along the circumference of the circle. 
In order to encode this construction in an unambiguous manner we work with signed spring lengths as degrees of freedom.
\sn{The orientation of a spring is chosen such that it goes from a node of lower index to a node of higher index.
This is an arbitrary choice, but defined orientations are essential in our formalism. 
The sign of the spring length is chosen to be positive if its orientation on the circle points counter-clockwise and negative otherwise.}

The network can be encoded within a \emph{graph representation}, where the springs \sn{together with their orientations} are the \emph{directed} edges of the graph, with signed lengths as edge weights (\cref{img:schematic}~(b)).
To lie on the circle, the graph and edge weights must be compatible in the sense that the sum of the edge weights around each cycle of the graph is equal to an integer, which we refer to as its winding number $g$.
Our network generation procedure guarantees this compatibility. It results in a \emph{random directed Hamiltonian graph}, i.e., 
a graph that contains a cycle that visits each node exactly once, with $N$ nodes and average degree $z$. This graph comes equipped with compatible initial spring lengths/edge weights $\{\bar l_i\}_{i=1}^{Nz/2}$ that are each \emph{uniformly} distributed as $\mathcal U(-0.5,0.5)$, but, since they are coupled by integer winding numbers, not mutually independent \cite{weiss2006course} as random variables.

We seek to characterize the length (i.e. force) distributions of springs in networks after they have relaxed into mechanical equilibrium. Relaxation preserves a network’s topology, i.e., its graph together with a set of winding numbers, that arises from the generation process. Note that networks sharing the same graph may have different sets of winding numbers, and therefore distinct relaxed states (\cref{img:schematic}~(c)).
A particular realization of an initial network (see above) uniquely determines network topology and results in a known linear solution operator for the respective mechanical equilibrium. However, a network ensemble realizes many topologies---yielding a random solution operator---making it more difficult to determine the ensemble-averaged distribution of relaxed lengths.

Motivated by experiments, where explicit information on particular realizations is hard to measure, we study ensembles with a fixed number of nodes $N$ and average degree $z$.
Surprisingly, such ensembles have characterizable force distributions despite varying topologies.
Explicitly accounting for these unknown underlying topologies makes our approach different from a mean-field description.


Formally, our setup can be written as the following optimization problem:
\begin{align}
  \text{minimize}\quad &\frac{1}{2}\boldsymbol l^T \boldsymbol l\quad \text{subject to}\quad \mathbf C \boldsymbol l = \boldsymbol g = \mathbf C \lbar\,,
  \label{eq:totalEnergy}
\end{align}
where $\boldsymbol l \in \mathbb R^{Nz/2}$ is the vector of all spring lengths and $\boldsymbol g \in \mathbb Z^m$ is the vector of winding numbers, which is determined by the vector of initial spring lengths $\lbar$ and the \emph{signed cycle matrix} $\mathbf C \in \mathbb Z^{m \times Nz/2}$\mw{, described below}.
\mw{Note that all the above quantities are random variables.}

The first part in \cref{eq:totalEnergy} minimizes the total elastic energy of the system, whereas the second part preserves the topology of the network by fixing the winding numbers of a set of $m=N(z/2-1)+1$ \emph{fundamental cycles} \footnote{A fundamental cycle is defined as a cycle that occurs when adding a single edge to a spanning tree of the graph. There are $N-1$ edges in the spanning tree, so $Nz/2 - (N-1)$ edges can be added. Hence, there are $N(z/2-1)+1$ fundamental cycles.}.
Note that the choice of fundamental cycles corresponds to a choice of basis and is therefore not unique.
The solution to \cref{eq:totalEnergy}, however, is independent of this choice \cite{heidemann2017a}.

After choosing a cycle basis, the $\mathbf C$-matrix is constructed by specifying an orientation for each fundamental cycle and then setting $C_{ji}$ equal to: $1$ if spring $i$ is part of the $j$th fundamental cycle and their orientations agree, or $-1$ if their orientations are opposite, and $0$ otherwise.
For the example in \cref{img:schematic}~(a), the cycle matrix and vector of winding numbers are given by $C_{1}=(1,1,1,-1,0,0)$, $C_{2}=(0,0,0,1,1,1)$, and $\boldsymbol g=(-1,0)^T$, respectively.
Note that winding numbers correspond to the signed number of times a cycle wraps around the circle. Contractible cycles have winding number zero.
If all cycles were contractible, then \cref{eq:totalEnergy} would have a trivial solution with all springs collapsed to a single point; it is only the presence of nontrivial cycle constraints that prevents this outcome.
\Cref{eq:totalEnergy} defines a \emph{quadratic programming problem} with a unique analytic solution.
Written in terms of the spring length changes $\dl$ during relaxation to the final configuration $\lstar$, the solution is:
\begin{align}
  \dl \coloneqq \lstar - \lbar = (\underbrace{\mathbf C^T(\mathbf C\mathbf C^T)^{-1}\mathbf C - \mathbf I)}_{\mathbf \eqqcolon \mathbf S} \lbar\,,
  \label{eq:kkt}
\end{align}
which can be explicitly computed for each realization via, e.g., the optimization library IPOPT \cite{Wachter2006}.

To express the resulting force distributions of our ensembles we consider the expected histogram of the vector $\lstar$ of random variables.
This results in a \emph{univariate} probability density $p_{\lstar}$, which is given by the average of the individual spring densities, i.e., $p_{\lstar}(\ellstar)\coloneqq\frac{2}{Nz} \sum_{i=1}^{Nz/2} p_{l^*_i}(\ellstar)$ (for details see~\cite{heidemann2017a}).
Using $l_i^* = \bar l_i + \Delta l_i$ (\cref{eq:kkt}) we compute for each component:
\begin{align}
  p_{l_i^*}(\ell^*) = \intinf p_{\bar l_i }(\bar \ell) \cdot p_{\Delta l_i|\bar l_i = \bar \ell}(\ell^* - \bar \ell)\, d\ellbar\,.
\end{align}
Since the initial spring lengths $\bar l_i$ are identically distributed, i.e., $p_{\bar l_i}=p_{\bar l}$, we obtain that
\begin{align}
  \label{eq:plf}
  p_{\lstar}(\ellstar) 
  = \intinf p_{\bar l}\, (\ellbar) \cdot p_{\dl|\lbar=\ellbar}\, (\ellstar-\ellbar) \,d\ellbar\,,\\
  \text{with}\quad p_{\dl | \lbar=\bar \ell}\,(\dell) \coloneqq \frac{2}{Nz} \sum_{i=1}^{Nz/2} p_{\Delta l_i | \bar l_i=\bar \ell}\,(\dell)\,.
  \label{eq:averagep}
\end{align}

In the following we characterize the conditional probability density \cref{eq:averagep} that completely determines the final distribution of spring lengths given the initial distribution (\cref{eq:plf}).
\Cref{eq:kkt} relates $\dl$ to $\lbar$ and a random matrix $\mathbf S$, which vary with the topology of each realization.
It is therefore challenging to obtain $p_{\dlgiven}$ explicitly, especially since the individual $\bar l_i$ are not mutually independent.
Instead, we consider the first two moments, $\E( \dlgiven )$ and $\vardlgivenl$, and investigate under which conditions $\dl|_{\lbar=\ellbar}$ is approximately normally distributed.

\Cref{eq:kkt,eq:averagep} lead to (see~\cite{heidemann2017a} for the derivation):
\begin{align}
  \E( \dlgiven )
  = \frac{2\,\ellbar}{Nz} \tr \mathbf S = - \frac{2\, \ellbar}{z} \left(1-\frac{1}{N}\right)\,, \label{eq:muconditional}
\end{align}
where  $\tr \mathbf S= 1-N$ is an invariant of the ensemble that surprisingly only depends on the number of nodes in the graphs, not on their respective topologies.
We compare \cref{eq:muconditional} to a \emph{mean-field} (mf) approach, 
where each node is displaced as if all other nodes in the network were fixed.
In this case ${\E( \dlgiven )|_{\text{mf}} = - 2 \ellbar/z }$ \cite{heidemann2017a}; in particular, the mean-field result agrees with the exact solution \cref{eq:muconditional} in the limit $N \to \infty$, i.e., there is no significant difference for large node numbers.
In contrast, we will show that for the variance, the mean-field solution differs substantially from the exact result, even in the limit $N \to \infty$.

The conditional variance $\vardlgivenl$ remains challenging to express analytically.
\mw{Indeed, in general, there are many graphs realizing the same $z$ and $N$, each with its own topology that may introduce nonzero covariance between the edge lengths.}


For two extreme cases, namely the cycle graph ($z=2$, $N>3$) and the complete graph ($z=N-1$, each node connected to every other node), there exists only a single possible graph, respectively, each being symmetric (i.e., vertex- and edge-transitive \cite{biggs1993algebraic}), allowing us to derive $\vardlgivenl$ explicitly.
\sn{In particular, edge transitivity allows us to reduce to a single entry in the $\lstar$ vector, which is given by a weighted sum of identically distributed, but dependent random variables (\cref{eq:kkt}).}

\sn{For the case of the cycle graph, an entry in \cref{eq:kkt} simplifies to $\Delta l_i = N^{-1}\sum_{j\neq i} \bar l_j$. Using conditional pairwise independence of the initial edge random variables allows for direct computation of $\vardlgivenl=(N-1)/N^2 \Var (\bar l)$.}
%

\sn{For the case of the complete graph, the derivation of the conditional variance is significantly more involved.
In order to obtain manageable algebraic expressions, {one} needs to carefully choose the cycle basis. 
This choice is detailed in \cite{heidemann2017a} and leads to {a} tractable analysis of $(\mathbf C \mathbf C^T)^{-1}$, which can then be applied to reformulate the problem in terms of conditionally independent winding number random variables, leading to
$\vardlgivenl = (N-2)/N^2 (|\ellbar| - \ellbar^2)$.}

\begin{figure*}
  \centering
  \includegraphics{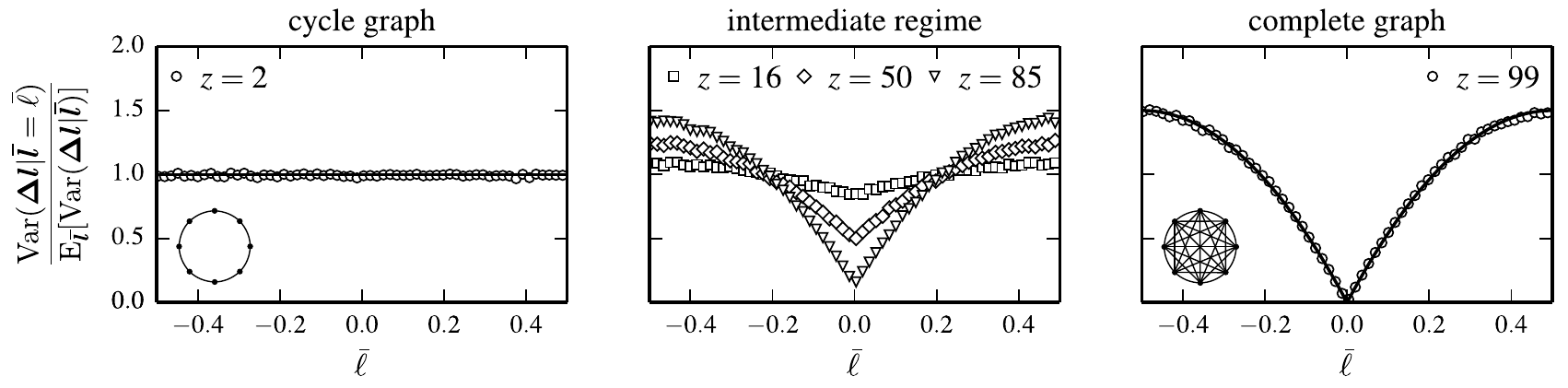}
  \caption{Normalized conditional variance $\vardlgivenl/\meanvar$ as a function of $\bar \ell$ for graphs with $N=100$ and varying $z$. For each value of $z$, data points correspond to ensemble averages (repeated simulations) with \num{4.95e6} springs in total. We use local linear regression with \num{3e4} nearest neighbors to estimate the variance for different values of $\bar \ell$. The solid lines correspond to the analytically derived expressions for the cycle and the complete graph (illustrated in the insets). In the intermediate regime of connectivity, the variance shows a continuous transition between the two extreme cases.}
  \label{fig:conditionalVariance}
\end{figure*}

\sn{For the intermediate-connectivity regime, $2<z<N-1$, a similar approach remains elusive; however, numerical data suggest that the conditional variance exhibits a continuous transition between the two extremes (\cref{fig:conditionalVariance}).}
We also observe that the conditional variance is approximately constant given that $z \ll N$. This is the most relevant case for biological networks where typically $z\lesssim4$.
For $z\ll N$, we may thus approximate $\vardlgivenl\approx\meanvar$\sn{, which we now derive.}

The \emph{law of total variance} \cite{weiss2006course} \sn{states}:
\begin{align}
\E_{\lbar}[\Var(\dl\,|\,\lbar)] = \Var( \dl ) - \Var_{\lbar}[\E( \dl\,|\,\lbar)]\,.
\label{eq:totalvar}
\end{align}
From \cref{eq:kkt,eq:averagep} it follows that
\begin{align}
  \Var( \dl ) = -\frac{2 \,\Var( \bar l )}{Nz} \tr \mathbf S
  =  \frac{2 }{z} \left(1-\frac{1}{N}\right)\Var( \bar l )\,,
  \label{eq:sigmadl2}
\end{align}
where we have used that $\mathbf S^2 = - \mathbf S$ (see~\cite{heidemann2017a} for details).
We can again compare this expression to its \emph{mean-field} counterpart: $\Var(\dl)|_{\text{mf}}= 2 /z (1+1/z))\varlbar$ (see~\cite{heidemann2017a}).
Clearly, the expressions do not agree in the limit $N\to\infty$. In particular, for sparsely connected networks (small values of $z$), there are significant deviations, independent of the number of nodes in the network.
Using \cref{eq:muconditional} we also have that
$\Var_{\lbar}[\E( \dl\,|\,\lbar)] = {(2/z(1-1/N))^2 \Var(\bar l)}$
and therefore by substituting into \cref{eq:totalvar}:
\begin{align}
  \frac{\E_{\lbar} [\vardlgiven]}{\varlbar} &=\frac{2}{z}\left(1-\frac{1}{N}\right)\left[1 - \frac{2}{z}\left(1-\frac{1}{N}\right)\right] \,.
  \label{eq:meanvar}
\end{align}

\begin{figure}
  \centering
  \includegraphics{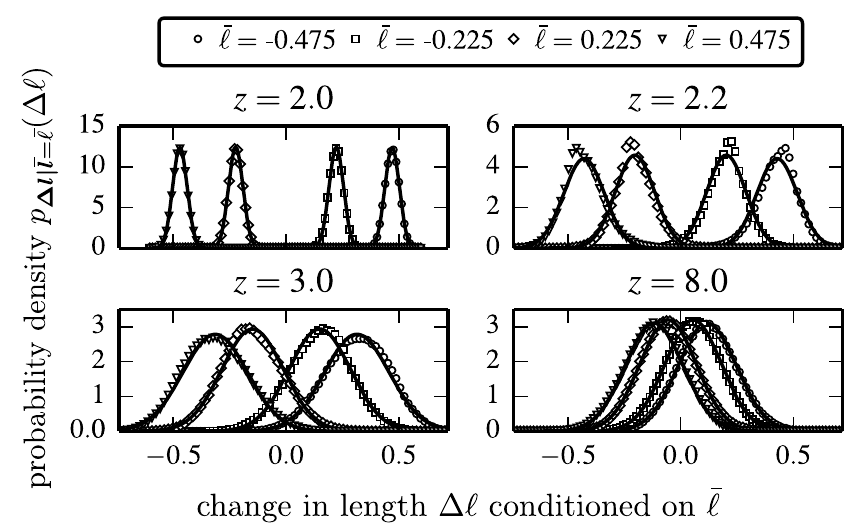}
  \caption{Conditional probability density $p_{\dlgiven}\,(\Delta \ell)$ for spring networks with $N=100$ and varying $z$, conditioned on different $\bar \ell$ values. For each value of $z$, data points correspond to ensemble averages (repeated simulations) with \num{4.95e6} springs in total. Solid lines correspond to best fit normal distributions. The cycle graph ($z=2$) is close to being normally distributed---as proven for $N\to\infty$. Whereas for $z=2.2$ there are still deviations from a normal distribution, for $z=3$ and larger the densities quickly become approximately normally distributed.}
  \label{fig:pdlconditional}
\end{figure}
Now, if $\dl|_{\lbar=\ellbar}$ were normally distributed, having estimates for mean and variance would be sufficient to fully characterize $p_{\dlgiven}$.
Indeed, for the two extremes, cycle and complete graph, we can prove that $\dl|_{\lbar=\ellbar} $ is \emph{normally distributed} in the limit $N \to \infty$, with a rate of convergence proportional to $(N-2)^{-1/2}$ \cite{heidemann2017a}.

This result might look like a direct application of the classical \emph{central limit theorem}.
\sn{However, since the edge lengths are not independent as random variables, more sophisticated techniques are required to represent the solution in terms of a suitable set of \sn{mutually} independent random variables.
In contrast to situations in time series analysis \cite{brockwell2009time}, where independence holds beyond a certain time window, the cycle constraints prohibit localization of dependencies.
\sn{To deal with this problem, we reduce the number of variables by relaxing each integer cycle constraint to an interval constraint.}
Harnessing the resulting independence then requires a non-standard transformation of random variables, which complicates a direct application of the Berry-Esseen theorem \cite{berry1941accuracy,esseen1942liapounoff} (a deviation-bound version of the central-limit theorem) to obtain a quantitative bound on the distance to a normal distribution.}

Recall that \sn{in the intermediate-connectivity regime, $2<z<N-1$, the ensembles contain graphs with varying cycle structures making a similar analysis significantly more challenging.}
\sn{In simulations}, \sn{however}, we observe that $\dl|_{\lbar=\ellbar}$ is approximately normally distributed if $z$ is sufficiently large (\cref{fig:pdlconditional}).

\begin{figure}
  \centering
  \includegraphics{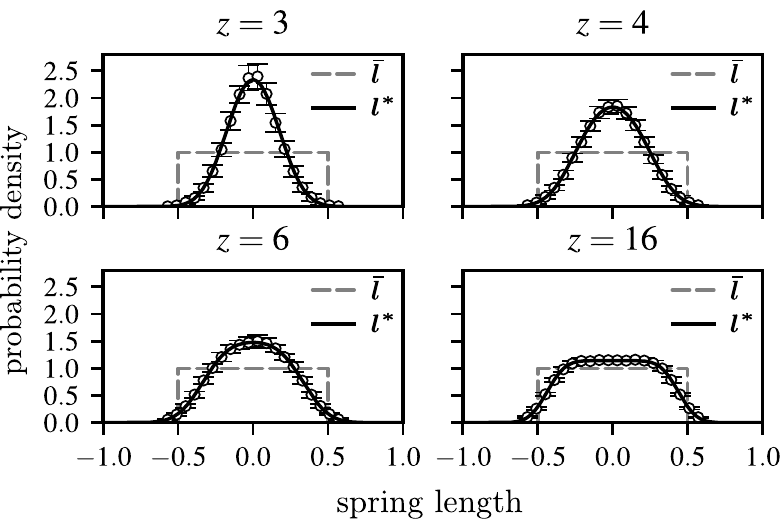}
  \caption{Probability density $p_{\lstar}(\ell^*)$ for the final spring lengths for networks with $N=1000$ and varying $z$. Solid black lines show the analytic expression for $p_{\lstar}(\ell^*)$ \cite{heidemann2017a}; data points correspond to averages over 50 simulations. The error bars correspond to the standard deviation. \sn{For comparison, we show the initial uniform spring length distribution $p_{\lbar}(\ellbar)$ as a gray dashed line.}}
  \label{fig:lengthdistribution}
\end{figure}

Our empirical observations and theoretical discussion above justify the following approximation for $3\leq z\ll N$:
\begin{align}
  \dl|_{\lbar=\ellbar} \sim \mathcal N\Big[\E( \dlgiven ),E_{\lbar}[\vardlgiven]\Big]\,,
  \label{eq:pdlconditional}
\end{align}
with the expressions for $\E( \dlgiven )$ and $E_{\lbar}[\vardlgiven]$ given in \cref{eq:muconditional,eq:meanvar}.
Using \cref{eq:plf,eq:pdlconditional} we obtain an explicit representation for the final length distribution $p_{\lstar}(\ellstar)$ in mechanical equilibrium (see~\cite{heidemann2017a}).
In \cref{fig:lengthdistribution} we compare this analytical expression to ensembles of simulated networks; we observe excellent agreement. 


In conclusion, we presented a probabilistic theory of force distributions in one-dimensional random spring networks on a circle.
Here we regarded networks with initially unbalanced forces that relax into mechanical equilibrium.
\sn{When drawing the analogy to a biological network, our approach, which focuses on the relaxation of the system after non-equilibrium starting conditions, is equivalent to assuming a separation of time scales where internal or external non-equilibrium processes slowly create forces in the network that rapidly equilibrate.}

We developed a \emph{graph-theoretical} approach that allows us to exactly compute mean and expected variance of the \mw{distribution of conditional length changes that completely determines the final length distribution.} 
For the two extreme cases, the \emph{cycle graph} and the \emph{complete graph}, we could additionally prove convergence of \sn{this distribution} to a \emph{normal distribution}.
A systematic analytical treatment of the---less symmetric---intermediate regime of connectivity is more demanding and not provided here.
\sn{However, our results suggest an approximation that shows excellent agreement with simulation for the biologically relevant regime of connectivity, $3\leq z\ll N$}.

It is straightforward to generalize the approach we presented here to higher spatial dimensions $d$ if the probability densities $p_{\lbar_k}$ for the components of the initial spring vectors are independent.
In that case, due to the linearity of spring forces with extension, the optimization problem decouples into the spatial components.
The probability density for the final spring vectors then is simply given as the product of the one-dimensional results:
\begin{align*}
  p_{\lstar}(\boldsymbol{\ell^*}) = \prod\limits_{k=1}^d p_{\lstar_k}(\ell^*_k)\,.
\end{align*}
Hence, our results carry over to two- and three-dimensional networks, which are more commonly studied in practice and are of biological and physiological relevance.

Interestingly, a classical \emph{mean-field} approach \emph{fails} to capture the mean and the variance of the relevant distributions. The error is particularly pronounced for the---biologically most relevant---regime of low degrees of connectivity, and does not vanish in the limit of infinite node number.
Our work demonstrates that \emph{network topology}---here manifested as \emph{cycle constraints}---is crucial for the correct determination of force distributions in an elastic spring network.

This opens the door for future research on the role that network topology plays in more complex elastic networks, e.g., in the presence of dynamics, spring nonlinearities or rupture.
Moreover, the mixture of probabilistic and graph-theoretical techniques may prove useful for other types of network theories.
\begin{acknowledgments}
  The authors would like to thank Friedrich Bös, Alexander Hartmann, and Fabian Telchow for fruitful discussions. Funding from the Deutsche Forschungsgemeinschaft (DFG) within the collaborative research center SFB 755, project A3, is gratefully acknowledged.
C.F.S was additionally supported by a European Research Council Advanced Grant PF7 ERC-2013-AdG, Project 340528.
\end{acknowledgments}

\bibliography{literature,references-mendeley}
\end{document}
%

%% file: shortcuts.tex
\newcommand{\iflanggerman}[2]{
 \iflanguage{german}{#1}{
  \iflanguage{ngerman}{#1}{#2}
 }
}

\iflanggerman{

}{

}

\DeclareMathOperator{\tr}{tr}

\newcommand{\intinf}{\int\limits_{-\infty}^{+\infty}}

\DeclareMathOperator\Var{Var}
\DeclareMathOperator\E{E}

%% file: circle-springs-curved-with-winding-N5.pdf_tex
\begingroup%
  \makeatletter%
  \providecommand\color[2][]{%
    \errmessage{(Inkscape) Color is used for the text in Inkscape, but the package 'color.sty' is not loaded}%
    \renewcommand\color[2][]{}%
  }%
  \providecommand\transparent[1]{%
    \errmessage{(Inkscape) Transparency is used (non-zero) for the text in Inkscape, but the package 'transparent.sty' is not loaded}%
    \renewcommand\transparent[1]{}%
  }%
  \providecommand\rotatebox[2]{#2}%
  \ifx\svgwidth\undefined%
    \setlength{\unitlength}{229.82626953bp}%
    \ifx\svgscale\undefined%
      \relax%
    \else%
      \setlength{\unitlength}{\unitlength * \real{\svgscale}}%
    \fi%
  \else%
    \setlength{\unitlength}{\svgwidth}%
  \fi%
  \global\let\svgwidth\undefined%
  \global\let\svgscale\undefined%
  \makeatother%
  \begin{picture}(1,1.00249637)%
    \put(0,0){\includegraphics[width=\unitlength]{circle-springs-curved-with-winding-N5.pdf}}%
    \put(-0.02613028,0.98713253){\color[rgb]{0,0,0}\makebox(0,0)[lb]{\smash{(a)}}}%
    \put(0.48403927,0.98711404){\color[rgb]{0,0,0}\makebox(0,0)[lb]{\smash{(b)}}}%
    \put(0.32184316,0.92486891){\color[rgb]{0,0,0}\makebox(0,0)[lb]{\smash{$l_1$}}}%
    \put(0.33208591,0.66036248){\color[rgb]{0,0,0}\makebox(0,0)[lb]{\smash{$l_2$}}}%
    \put(0.04998317,0.65854682){\color[rgb]{0,0,0}\makebox(0,0)[lb]{\smash{$l_3$}}}%
    \put(0.0745651,0.94688877){\color[rgb]{0,0,0}\makebox(0,0)[lb]{\smash{$l_4$}}}%
    \put(0.07884971,0.84120963){\color[rgb]{0,0,0}\makebox(0,0)[lb]{\smash{$l_5$}}}%
    \put(0.13235554,0.89513836){\color[rgb]{0,0,0}\makebox(0,0)[lb]{\smash{$l_6$}}}%
    \put(0.35034603,0.37229672){\color[rgb]{0,0,0}\makebox(0,0)[lb]{\smash{$g=0$}}}%
    \put(0.56727077,0.37229672){\color[rgb]{0,0,0}\makebox(0,0)[lb]{\smash{$g=-1$}}}%
    \put(0.81815162,0.36964016){\color[rgb]{0,0,0}\makebox(0,0)[lb]{\smash{$g=2$}}}%
    \put(-0.02494674,0.49658196){\color[rgb]{0,0,0}\makebox(0,0)[lb]{\smash{(c)}}}%
    \put(0.2393679,0.12455864){\color[rgb]{0,0,0}\rotatebox{90}{\makebox(0,0)[b]{\smash{relaxed}}}}%
    \put(0.2393679,0.38023538){\color[rgb]{0,0,0}\rotatebox{90}{\makebox(0,0)[b]{\smash{initial}}}}%
    \put(0.35034603,0.11614281){\color[rgb]{0,0,0}\makebox(0,0)[lb]{\smash{$g=0$}}}%
    \put(0.56727077,0.11614281){\color[rgb]{0,0,0}\makebox(0,0)[lb]{\smash{$g=-1$}}}%
    \put(0.81815162,0.11614281){\color[rgb]{0,0,0}\makebox(0,0)[lb]{\smash{$g=2$}}}%
    \put(0.82262351,0.97010476){\color[rgb]{0,0,0}\makebox(0,0)[lb]{\smash{$l_1$}}}%
    \put(0.99220696,0.78799774){\color[rgb]{0,0,0}\makebox(0,0)[lb]{\smash{$l_2$}}}%
    \put(0.81804252,0.62674482){\color[rgb]{0,0,0}\makebox(0,0)[lb]{\smash{$l_3$}}}%
    \put(0.72192839,0.86976041){\color[rgb]{0,0,0}\makebox(0,0)[lb]{\smash{$l_4$}}}%
    \put(0.55919072,0.70661642){\color[rgb]{0,0,0}\makebox(0,0)[lb]{\smash{$l_5$}}}%
    \put(0.57801003,0.92563864){\color[rgb]{0,0,0}\makebox(0,0)[lb]{\smash{$l_6$}}}%
    \put(0.08122487,0.40608625){\color[rgb]{0,0,0}\makebox(0,0)[lb]{\smash{cycle}}}%
    \put(0.07472858,0.37133259){\color[rgb]{0,0,0}\makebox(0,0)[lb]{\smash{graph}}}%
    \put(0.06399543,0.11474909){\color[rgb]{0,0,0}\makebox(0,0)[lb]{\smash{$N=5$}}}%
    \put(0.07970559,0.07227178){\color[rgb]{0,0,0}\makebox(0,0)[lb]{\smash{$z=2$}}}%
    \put(0.41257225,0.52132339){\color[rgb]{0,0,0}\makebox(0,0)[lb]{\smash{varying winding numbers}}}%
  \end{picture}%
\endgroup%

%% file: spring-networks-letter.bbl
\begin{thebibliography}{21}%
\makeatletter
\providecommand \@ifxundefined [1]{%
 \@ifx{#1\undefined}
}%
\providecommand \@ifnum [1]{%
 \ifnum #1\expandafter \@firstoftwo
 \else \expandafter \@secondoftwo
 \fi
}%
\providecommand \@ifx [1]{%
 \ifx #1\expandafter \@firstoftwo
 \else \expandafter \@secondoftwo
 \fi
}%
\providecommand \natexlab [1]{#1}%
\providecommand \enquote  [1]{``#1''}%
\providecommand \bibnamefont  [1]{#1}%
\providecommand \bibfnamefont [1]{#1}%
\providecommand \citenamefont [1]{#1}%
\providecommand \href@noop [0]{\@secondoftwo}%
\providecommand \href [0]{\begingroup \@sanitize@url \@href}%
\providecommand \@href[1]{\@@startlink{#1}\@@href}%
\providecommand \@@href[1]{\endgroup#1\@@endlink}%
\providecommand \@sanitize@url [0]{\catcode `\\12\catcode `\$12\catcode
  `\&12\catcode `\#12\catcode `\^12\catcode `\_12\catcode `\%12\relax}%
\providecommand \@@startlink[1]{}%
\providecommand \@@endlink[0]{}%
\providecommand \url  [0]{\begingroup\@sanitize@url \@url }%
\providecommand \@url [1]{\endgroup\@href {#1}{\urlprefix }}%
\providecommand \urlprefix  [0]{URL }%
\providecommand \Eprint [0]{\href }%
\providecommand \doibase [0]{http://dx.doi.org/}%
\providecommand \selectlanguage [0]{\@gobble}%
\providecommand \bibinfo  [0]{\@secondoftwo}%
\providecommand \bibfield  [0]{\@secondoftwo}%
\providecommand \translation [1]{[#1]}%
\providecommand \BibitemOpen [0]{}%
\providecommand \bibitemStop [0]{}%
\providecommand \bibitemNoStop [0]{.\EOS\space}%
\providecommand \EOS [0]{\spacefactor3000\relax}%
\providecommand \BibitemShut  [1]{\csname bibitem#1\endcsname}%
\let\auto@bib@innerbib\@empty
\bibitem [{\citenamefont {Heussinger}\ and\ \citenamefont
  {Frey}(2007)}]{heussinger2007force}%
  \BibitemOpen
  \bibfield  {author} {\bibinfo {author} {\bibfnamefont {C.}~\bibnamefont
  {Heussinger}}\ and\ \bibinfo {author} {\bibfnamefont {E.}~\bibnamefont
  {Frey}},\ }\href {\doibase 10.1140/epje/i2007-10209-1} {\bibfield  {journal}
  {\bibinfo  {journal} {Eur. Phys. J. E}\ }\textbf {\bibinfo {volume} {24}},\
  \bibinfo {pages} {47} (\bibinfo {year} {2007})}\BibitemShut {NoStop}%
\bibitem [{\citenamefont {Arevalo}\ \emph {et~al.}(2015)\citenamefont
  {Arevalo}, \citenamefont {Kumar}, \citenamefont {Urbach},\ and\ \citenamefont
  {Blair}}]{arevalo2015stress}%
  \BibitemOpen
  \bibfield  {author} {\bibinfo {author} {\bibfnamefont {R.~C.}\ \bibnamefont
  {Arevalo}}, \bibinfo {author} {\bibfnamefont {P.}~\bibnamefont {Kumar}},
  \bibinfo {author} {\bibfnamefont {J.~S.}\ \bibnamefont {Urbach}}, \ and\
  \bibinfo {author} {\bibfnamefont {D.~L.}\ \bibnamefont {Blair}},\ }\href
  {\doibase 10.1371/journal.pone.0118021} {\bibfield  {journal} {\bibinfo
  {journal} {PloS ONE}\ }\textbf {\bibinfo {volume} {10}},\ \bibinfo {pages}
  {e0118021} (\bibinfo {year} {2015})}\BibitemShut {NoStop}%
\bibitem [{\citenamefont {Mizuno}\ \emph {et~al.}(2007)\citenamefont {Mizuno},
  \citenamefont {Tardin}, \citenamefont {Schmidt},\ and\ \citenamefont
  {MacKintosh}}]{mizuno2007nonequilibrium}%
  \BibitemOpen
  \bibfield  {author} {\bibinfo {author} {\bibfnamefont {D.}~\bibnamefont
  {Mizuno}}, \bibinfo {author} {\bibfnamefont {C.}~\bibnamefont {Tardin}},
  \bibinfo {author} {\bibfnamefont {C.~F.}\ \bibnamefont {Schmidt}}, \ and\
  \bibinfo {author} {\bibfnamefont {F.~C.}\ \bibnamefont {MacKintosh}},\ }\href
  {\doibase 10.1126/science.1134404} {\bibfield  {journal} {\bibinfo  {journal}
  {Science}\ }\textbf {\bibinfo {volume} {315}},\ \bibinfo {pages} {370}
  (\bibinfo {year} {2007})}\BibitemShut {NoStop}%
\bibitem [{\citenamefont {Koenderink}\ \emph {et~al.}(2009)\citenamefont
  {Koenderink}, \citenamefont {Dogic}, \citenamefont {Nakamura}, \citenamefont
  {Bendix}, \citenamefont {MacKintosh}, \citenamefont {Hartwig}, \citenamefont
  {Stossel},\ and\ \citenamefont {Weitz}}]{koenderink2009active}%
  \BibitemOpen
  \bibfield  {author} {\bibinfo {author} {\bibfnamefont {G.~H.}\ \bibnamefont
  {Koenderink}}, \bibinfo {author} {\bibfnamefont {Z.}~\bibnamefont {Dogic}},
  \bibinfo {author} {\bibfnamefont {F.}~\bibnamefont {Nakamura}}, \bibinfo
  {author} {\bibfnamefont {P.~M.}\ \bibnamefont {Bendix}}, \bibinfo {author}
  {\bibfnamefont {F.~C.}\ \bibnamefont {MacKintosh}}, \bibinfo {author}
  {\bibfnamefont {J.~H.}\ \bibnamefont {Hartwig}}, \bibinfo {author}
  {\bibfnamefont {T.~P.}\ \bibnamefont {Stossel}}, \ and\ \bibinfo {author}
  {\bibfnamefont {D.~A.}\ \bibnamefont {Weitz}},\ }\href {\doibase
  10.1073/pnas.0903974106} {\bibfield  {journal} {\bibinfo  {journal} {Proc.
  Natl. Acad. Sci. USA}\ }\textbf {\bibinfo {volume} {106}},\ \bibinfo {pages}
  {15192} (\bibinfo {year} {2009})}\BibitemShut {NoStop}%
\bibitem [{\citenamefont {Alvarado}\ \emph {et~al.}(2013)\citenamefont
  {Alvarado}, \citenamefont {Sheinman}, \citenamefont {Sharma}, \citenamefont
  {MacKintosh},\ and\ \citenamefont {Koenderink}}]{alvarado2013molecular}%
  \BibitemOpen
  \bibfield  {author} {\bibinfo {author} {\bibfnamefont {J.}~\bibnamefont
  {Alvarado}}, \bibinfo {author} {\bibfnamefont {M.}~\bibnamefont {Sheinman}},
  \bibinfo {author} {\bibfnamefont {A.}~\bibnamefont {Sharma}}, \bibinfo
  {author} {\bibfnamefont {F.~C.}\ \bibnamefont {MacKintosh}}, \ and\ \bibinfo
  {author} {\bibfnamefont {G.~H.}\ \bibnamefont {Koenderink}},\ }\href
  {\doibase 10.1038/nphys2715} {\bibfield  {journal} {\bibinfo  {journal} {Nat.
  Phys.}\ }\textbf {\bibinfo {volume} {9}},\ \bibinfo {pages} {591} (\bibinfo
  {year} {2013})}\BibitemShut {NoStop}%
\bibitem [{\citenamefont {Shah}\ and\ \citenamefont
  {Janmey}(1997)}]{shah1997strain}%
  \BibitemOpen
  \bibfield  {author} {\bibinfo {author} {\bibfnamefont {J.~V.}\ \bibnamefont
  {Shah}}\ and\ \bibinfo {author} {\bibfnamefont {P.~A.}\ \bibnamefont
  {Janmey}},\ }\href {\doibase 10.1007/BF00366667} {\bibfield  {journal}
  {\bibinfo  {journal} {Rheol. Acta}\ }\textbf {\bibinfo {volume} {36}},\
  \bibinfo {pages} {262} (\bibinfo {year} {1997})}\BibitemShut {NoStop}%
\bibitem [{\citenamefont {Lam}\ \emph {et~al.}(2011)\citenamefont {Lam},
  \citenamefont {Chaudhuri}, \citenamefont {Crow}, \citenamefont {Webster},
  \citenamefont {Li}, \citenamefont {Kita}, \citenamefont {Huang},\ and\
  \citenamefont {Fletcher}}]{lam2011mechanics}%
  \BibitemOpen
  \bibfield  {author} {\bibinfo {author} {\bibfnamefont {W.~A.}\ \bibnamefont
  {Lam}}, \bibinfo {author} {\bibfnamefont {O.}~\bibnamefont {Chaudhuri}},
  \bibinfo {author} {\bibfnamefont {A.}~\bibnamefont {Crow}}, \bibinfo {author}
  {\bibfnamefont {K.~D.}\ \bibnamefont {Webster}}, \bibinfo {author}
  {\bibfnamefont {T.-D.}\ \bibnamefont {Li}}, \bibinfo {author} {\bibfnamefont
  {A.}~\bibnamefont {Kita}}, \bibinfo {author} {\bibfnamefont {J.}~\bibnamefont
  {Huang}}, \ and\ \bibinfo {author} {\bibfnamefont {D.~A.}\ \bibnamefont
  {Fletcher}},\ }\href {\doibase 10.1038/nmat2903} {\bibfield  {journal}
  {\bibinfo  {journal} {Nat. Mater.}\ }\textbf {\bibinfo {volume} {10}},\
  \bibinfo {pages} {61} (\bibinfo {year} {2011})}\BibitemShut {NoStop}%
\bibitem [{\citenamefont {Heidemann}\ \emph {et~al.}(2015)\citenamefont
  {Heidemann}, \citenamefont {Sharma}, \citenamefont {Rehfeldt}, \citenamefont
  {Schmidt},\ and\ \citenamefont {Wardetzky}}]{Heidemann2014}%
  \BibitemOpen
  \bibfield  {author} {\bibinfo {author} {\bibfnamefont {K.~M.}\ \bibnamefont
  {Heidemann}}, \bibinfo {author} {\bibfnamefont {A.}~\bibnamefont {Sharma}},
  \bibinfo {author} {\bibfnamefont {F.}~\bibnamefont {Rehfeldt}}, \bibinfo
  {author} {\bibfnamefont {C.~F.}\ \bibnamefont {Schmidt}}, \ and\ \bibinfo
  {author} {\bibfnamefont {M.}~\bibnamefont {Wardetzky}},\ }\href {\doibase
  10.1039/C4SM01789G} {\bibfield  {journal} {\bibinfo  {journal} {Soft Matter}\
  }\textbf {\bibinfo {volume} {11}},\ \bibinfo {pages} {343} (\bibinfo {year}
  {2015})}\BibitemShut {NoStop}%
\bibitem [{\citenamefont {Feng}\ \emph {et~al.}(1985)\citenamefont {Feng},
  \citenamefont {Thorpe},\ and\ \citenamefont {Garboczi}}]{Thorpe1985}%
  \BibitemOpen
  \bibfield  {author} {\bibinfo {author} {\bibfnamefont {S.}~\bibnamefont
  {Feng}}, \bibinfo {author} {\bibfnamefont {M.~F.}\ \bibnamefont {Thorpe}}, \
  and\ \bibinfo {author} {\bibfnamefont {E.}~\bibnamefont {Garboczi}},\ }\href
  {\doibase 10.1103/PhysRevB.31.276} {\bibfield  {journal} {\bibinfo  {journal}
  {Phys. Rev. B}\ }\textbf {\bibinfo {volume} {31}},\ \bibinfo {pages} {276}
  (\bibinfo {year} {1985})}\BibitemShut {NoStop}%
\bibitem [{\citenamefont {Broedersz}\ \emph {et~al.}(2011)\citenamefont
  {Broedersz}, \citenamefont {Mao}, \citenamefont {Lubensky},\ and\
  \citenamefont {MacKintosh}}]{Broedersz2011b}%
  \BibitemOpen
  \bibfield  {author} {\bibinfo {author} {\bibfnamefont {C.~P.}\ \bibnamefont
  {Broedersz}}, \bibinfo {author} {\bibfnamefont {X.}~\bibnamefont {Mao}},
  \bibinfo {author} {\bibfnamefont {T.~C.}\ \bibnamefont {Lubensky}}, \ and\
  \bibinfo {author} {\bibfnamefont {F.~C.}\ \bibnamefont {MacKintosh}},\ }\href
  {\doibase 10.1038/nphys2127} {\bibfield  {journal} {\bibinfo  {journal} {Nat.
  Phys.}\ }\textbf {\bibinfo {volume} {7}},\ \bibinfo {pages} {983} (\bibinfo
  {year} {2011})}\BibitemShut {NoStop}%
\bibitem [{\citenamefont {Sheinman}\ \emph {et~al.}(2012)\citenamefont
  {Sheinman}, \citenamefont {Broedersz},\ and\ \citenamefont
  {MacKintosh}}]{Sheinman2012a}%
  \BibitemOpen
  \bibfield  {author} {\bibinfo {author} {\bibfnamefont {M.}~\bibnamefont
  {Sheinman}}, \bibinfo {author} {\bibfnamefont {C.~P.}\ \bibnamefont
  {Broedersz}}, \ and\ \bibinfo {author} {\bibfnamefont {F.~C.}\ \bibnamefont
  {MacKintosh}},\ }\href {\doibase 10.1103/PhysRevE.85.021801} {\bibfield
  {journal} {\bibinfo  {journal} {Phys. Rev. E}\ }\textbf {\bibinfo {volume}
  {85}},\ \bibinfo {pages} {021801} (\bibinfo {year} {2012})}\BibitemShut
  {NoStop}%
\bibitem [{\citenamefont {Storm}\ \emph {et~al.}(2005)\citenamefont {Storm},
  \citenamefont {Pastore}, \citenamefont {Mackintosh}, \citenamefont
  {Lubensky},\ and\ \citenamefont {Janmey}}]{Storm2005}%
  \BibitemOpen
  \bibfield  {author} {\bibinfo {author} {\bibfnamefont {C.}~\bibnamefont
  {Storm}}, \bibinfo {author} {\bibfnamefont {J.~J.}\ \bibnamefont {Pastore}},
  \bibinfo {author} {\bibfnamefont {F.~C.}\ \bibnamefont {Mackintosh}},
  \bibinfo {author} {\bibfnamefont {T.~C.}\ \bibnamefont {Lubensky}}, \ and\
  \bibinfo {author} {\bibfnamefont {P.~A.}\ \bibnamefont {Janmey}},\ }\href
  {\doibase 10.1038/nature03497.1.} {\bibfield  {journal} {\bibinfo  {journal}
  {Nature}\ }\textbf {\bibinfo {volume} {435}},\ \bibinfo {pages} {191}
  (\bibinfo {year} {2005})}\BibitemShut {NoStop}%
\bibitem [{\citenamefont {Sharma}\ \emph {et~al.}(2013)\citenamefont {Sharma},
  \citenamefont {Sheinman}, \citenamefont {Heidemann},\ and\ \citenamefont
  {MacKintosh}}]{Sharma2013a}%
  \BibitemOpen
  \bibfield  {author} {\bibinfo {author} {\bibfnamefont {A.}~\bibnamefont
  {Sharma}}, \bibinfo {author} {\bibfnamefont {M.}~\bibnamefont {Sheinman}},
  \bibinfo {author} {\bibfnamefont {K.~M.}\ \bibnamefont {Heidemann}}, \ and\
  \bibinfo {author} {\bibfnamefont {F.~C.}\ \bibnamefont {MacKintosh}},\ }\href
  {\doibase 10.1103/PhysRevE.88.052705} {\bibfield  {journal} {\bibinfo
  {journal} {Phys. Rev. E}\ }\textbf {\bibinfo {volume} {88}},\ \bibinfo
  {pages} {052705} (\bibinfo {year} {2013})}\BibitemShut {NoStop}%
\bibitem [{\citenamefont {Weiss}(2006)}]{weiss2006course}%
  \BibitemOpen
  \bibfield  {author} {\bibinfo {author} {\bibfnamefont {N.~A.}\ \bibnamefont
  {Weiss}},\ }\href
  {https://books.google.de/books/about/A_Course_in_Probability.html?id=p-rwJAAACAAJ&redir_esc=y}
  {\emph {\bibinfo {title} {A course in probability}}}\ (\bibinfo  {publisher}
  {Pearson},\ \bibinfo {year} {2006})\BibitemShut {NoStop}%
\bibitem [{Note1()}]{Note1}%
  \BibitemOpen
  \bibinfo {note} {A fundamental cycle is defined as a cycle that occurs when
  adding a single edge to a spanning tree of the graph. There are $N-1$ edges
  in the spanning tree, so $Nz/2 - (N-1)$ edges can be added. Hence, there are
  $N(z/2-1)+1$ fundamental cycles.}\BibitemShut {Stop}%
\bibitem [{\citenamefont {Heidemann}\ \emph {et~al.}()\citenamefont
  {Heidemann}, \citenamefont {Sageman-Furnas}, \citenamefont {Sharma},
  \citenamefont {Rehfeldt}, \citenamefont {Schmidt},\ and\ \citenamefont
  {Wardetzky}}]{heidemann2017a}%
  \BibitemOpen
  \bibfield  {author} {\bibinfo {author} {\bibfnamefont {K.~M.}\ \bibnamefont
  {Heidemann}}, \bibinfo {author} {\bibfnamefont {A.~O.}\ \bibnamefont
  {Sageman-Furnas}}, \bibinfo {author} {\bibfnamefont {A.}~\bibnamefont
  {Sharma}}, \bibinfo {author} {\bibfnamefont {F.}~\bibnamefont {Rehfeldt}},
  \bibinfo {author} {\bibfnamefont {C.~F.}\ \bibnamefont {Schmidt}}, \ and\
  \bibinfo {author} {\bibfnamefont {M.}~\bibnamefont {Wardetzky}},\ }\href@noop
  {} {\ }\Eprint {http://arxiv.org/abs/1707.01549} {arXiv:1707.01549}
  \BibitemShut {NoStop}%
\bibitem [{\citenamefont {W{\"{a}}chter}\ and\ \citenamefont
  {Biegler}(2005)}]{Wachter2006}%
  \BibitemOpen
  \bibfield  {author} {\bibinfo {author} {\bibfnamefont {A.}~\bibnamefont
  {W{\"{a}}chter}}\ and\ \bibinfo {author} {\bibfnamefont {L.~T.}\ \bibnamefont
  {Biegler}},\ }\href {\doibase 10.1007/s10107-004-0559-y} {\bibfield
  {journal} {\bibinfo  {journal} {Math. Program.}\ }\textbf {\bibinfo {volume}
  {106}},\ \bibinfo {pages} {25} (\bibinfo {year} {2005})}\BibitemShut
  {NoStop}%
\bibitem [{\citenamefont {Biggs}(1993)}]{biggs1993algebraic}%
  \BibitemOpen
  \bibfield  {author} {\bibinfo {author} {\bibfnamefont {N.}~\bibnamefont
  {Biggs}},\ }\href
  {https://books.google.de/books/about/Algebraic_Graph_Theory.html?id=6TasRmIFOxQC&redir_esc=y}
  {\emph {\bibinfo {title} {Algebraic Graph Theory}}}\ (\bibinfo  {publisher}
  {CUP},\ \bibinfo {address} {Cambridge, England},\ \bibinfo {year}
  {1993})\BibitemShut {NoStop}%
\bibitem [{\citenamefont {Brockwell}\ and\ \citenamefont
  {Davis}(2009)}]{brockwell2009time}%
  \BibitemOpen
  \bibfield  {author} {\bibinfo {author} {\bibfnamefont {P.}~\bibnamefont
  {Brockwell}}\ and\ \bibinfo {author} {\bibfnamefont {R.}~\bibnamefont
  {Davis}},\ }\href {http://www.springer.com/de/book/9780387974293} {\emph
  {\bibinfo {title} {Time Series: Theory and Methods}}},\ SSS\ (\bibinfo
  {publisher} {Springer},\ \bibinfo {address} {New York, NY},\ \bibinfo {year}
  {2009})\BibitemShut {NoStop}%
\bibitem [{\citenamefont {Berry}(1941)}]{berry1941accuracy}%
  \BibitemOpen
  \bibfield  {author} {\bibinfo {author} {\bibfnamefont {A.~C.}\ \bibnamefont
  {Berry}},\ }\href {\doibase 10.1090/S0002-9947-1941-0003498-3} {\bibfield
  {journal} {\bibinfo  {journal} {Trans. Amer. Math. Soc.}\ }\textbf {\bibinfo
  {volume} {49}},\ \bibinfo {pages} {122} (\bibinfo {year} {1941})}\BibitemShut
  {NoStop}%
\bibitem [{\citenamefont {Esseen}(1942)}]{esseen1942liapounoff}%
  \BibitemOpen
  \bibfield  {author} {\bibinfo {author} {\bibfnamefont {C.-G.}\ \bibnamefont
  {Esseen}},\ }\href@noop {} {\bibfield  {journal} {\bibinfo  {journal} {Ark.
  Mat. Astr. Fys.}\ }\textbf {\bibinfo {volume} {A28}},\ \bibinfo {pages} {1}
  (\bibinfo {year} {1942})}\BibitemShut {NoStop}%
\end{thebibliography}%
